

\documentstyle{amsppt}
\magnification\magstep1
\TagsOnRight
\baselineskip=12pt 
\font\ninepoint=cmr9
\font\titulo=cmbx10 scaled\magstep2
\def\1{\'{\i}}                         

\NoBlackBoxes
\def\ee{\varepsilon}
\def\calI{\Cal I}
\def\calS{\Cal S}
\def\aa#1{\alpha_{#1}\,}
\def\bb#1#2{\beta_{#1}^{\,\,#2}\,}
\def\cc#1{\gamma_{#1}\,}
\def\eee#1#2{\ee_{
{\underset{\rightarrow}\to{#1}},{\underset{\rightarrow}\to{#2}}}}
\def\k{\kappa}
\def\lra{\leftrightarrow}

\def\IW{1}
\def\Salet{2}
\def\Gilmor{3}
\def\MonPat{4}
\def\MooPat{5}
\def\PatZas{6}
\def\MonPatTol{7}
\def\COS{8}
\def\YasYagRos{9}
\def\Ros{10}
\def\Torret{11}
\def\Sommer{12}
\def\Herranz{13}
\def\SHO{14}
\def\HOS{15}
\def\BHOSi{16}
\def\BHOSii{17}
\def\BerMon{18}

\
\vskip 3cm

\centerline {\titulo CAYLEY--KLEIN ALGEBRAS AS}
\bigskip
\centerline {\titulo GRADED CONTRACTIONS OF {SO(N+1)}}
\bigskip

\vskip 1.5cm
\centerline {F.J. HERRANZ$^1$, M. DE MONTIGNY$^2$,
  M.A. DEL OLMO$^1$ and M. SANTANDER$^1$ }

\bigskip
\medskip

\centerline {\it 1. Departamento de F\'\i sica Te\'orica,
Universidad de Valladolid,}

\centerline{\it E--47011 Valladolid, Spain.}

\smallskip
\centerline{e-mail: fteorica\@cpd.uva.es}
\bigskip
\medskip

\centerline {\it 2. D\'epartement de Physique et Centre de
Recherches Math\'ematiques,}

\centerline {\it Universit\'e de Montr\'eal, C.P. 6128-A,}

\centerline {\it Montreal, (Quebec) H3C 3J7, Canada}

\centerline {\it (Present address: Department of Physics, High
 Energy Theory,}

\centerline {\it McGill University,}

\centerline {\it Montreal, (Quebec) H3A 2T8, Canada)}

\vskip 2truecm

\ninepoint
\noindent {{\bf Abstract.}
We study $\Bbb Z_2^{\otimes N}$ graded contractions of
the real compact simple Lie algebra $so(N+1)$, and we identify within
them the Cayley--Klein algebras as a naturally distinguished
subset.}
\rm
\vskip 2cm

\centerline {Revised Version: 10 Dec 93}

\bigskip
\bigskip

\baselineskip=12pt 

\vfill\eject

\centerline {\bf I. INTRODUCTION.}
\medskip
\medskip

The idea of group contraction was first explicitly introduced by
In\"on\"u and Wigner \cite{\IW}, in relation with the study of the
non--relativistic limit. In spite of a considerable body of literature
(see \cite{\Salet,\Gilmor} and references therein), contractions, both
of groups and especially of representations have remained an area full
of difficulties. Recently, de Montigny and Patera \cite {\MonPat}, and
Moody and Patera \cite{\MooPat} have developed a more general formalism
of contractions making use of the theory of Lie gradings. The crucial
property of graded contractions is that they preserve  a chosen grading
\cite{\PatZas} of the Lie algebra $L$ to be contracted. These ``graded
contractions'' contain as particular cases the In\"on\"u-Wigner
contractions, but they go well beyond these cases, and embrace into a
unified framework the study of algebra contractions and their
finite--dimensional representations.

\medskip

A nice feature of this approach is that the equations that determine all
possible contractions of a Lie algebra $L$ compatible with a given
grading, can be solved at once for all algebras admitting the same
grading. Reference \cite{\MonPat} contains tables for the grading groups
$\Bbb Z_2,\ \Bbb Z_3$ and $\Bbb Z_2\otimes\Bbb Z_2$.  An
interesting application of this theory \cite{\MonPatTol} is the fact
that a family of $\Bbb Z_2\otimes\Bbb Z_2$ graded contractions of the
algebra $so(5)$ (or of any real form of the complex Lie algebra $B_2$)
contains the kinematical algebras. (See \cite{\COS} where an
implicit use of gradings is also made in relation to kinematical
groups.)

\medskip

In this paper we show how the classical family of (orthogonal)
Cayley--Klein groups fits in a very natural way within the family of all
$\Bbb Z_2^{\otimes N}$ graded contractions of $so(N+1)$. The name
Cayley--Klein (CK) is linked with the appearance of these groups within
the context of Klein's consideration of most geometries as subgeometries
of projective geometry and to Cayley's theory of projective metrics
(e.g. see \cite{\YasYagRos,\Ros}). However, the complete
classification of these systems was not given by Klein himself. The
$N=2$ case was studied under the name of ``quadratic geometries" by
Poincar\'e, using essentially a modern group theoretical approach
(e.g. see \cite{\Torret}), and the classification for arbitrary
dimension $N$ was given in 1910 by Sommerville \cite{\Sommer}, who
showed that there are $3^N$ different systems in dimension $N$, each
corresponding to a choice of the kind of measure of distance between
points, lines, \dots , hyperplanes being either elliptic, parabolic
or hyperbolic.

\medskip

{}From a modern point of view, these geometries can be looked at as
systems of interlinked sets of symmetric homogeneous spaces
associated to a group $G$. In the general description for arbitrary
$N$ \cite{\Herranz--\HOS} (see also \cite{\BHOSi,\BHOSii} for
$N=2,3$), a CK geometry is a system characterized by an $\frac
{N(N+1)}2$--dimensional real Lie group $G$ and a set of $N$ basic
commuting involutions $S^{(k)}$ of the corresponding Lie algebra
$\frak g$. The generators of $\frak g$ invariant under the
involutions  ($S^{(0)}$, $S^{(1)}$,\dots, $S^{(N-1)}$) span $N$
subgroups of $G$ ($H^{(0)}$, $H^{(1)}$,\dots, $H^{(N-1)}$) which are
considered as isotopy groups of a point, line,\dots, hyperplane (or
$(N-1)$--flat) so that the $N$ symmetric homogeneous spaces
($G/H^{(0)}$, $G/H^{(1)}$,\dots, $G/H^{(N-1)}$) are the spaces of
points, lines,\dots, hyperplanes. The CK Lie group $G$ itself is
determined by $N$ fundamental real parameters
$(\kappa_1,\kappa_2,\dots,\kappa_N)$. Denoting by $J_{ij}\
(i,j=0,\dots,N; i<j)$ a basis of the CK Lie algebra $\frak
g_{(\kappa_1,\dots,\kappa_N)}$, we have the commutation relations
\cite{\Herranz--\HOS}
$$
[J_{ij},J_{lm}]
=\delta_{im}J_{lj}-\delta_{jl}J_{im}+\delta_{jm}\k_{lm}J_{il}+
\delta_{il}\k_{ij}J_{jm},\tag1.1
$$
where $i\leq l,\ j\leq m$, and the coefficients $\k_{ij}$ are defined by
$ \k_{ij}= \prod_{m=i+1}^j\k_m$, $ i,j=0,1,\dots,N;\  (i<j) $. Each
parameter $\k_1, \k_2, \dots ,\k_{N}$ is related to the kind of measure
between two points, lines, \dots , hyperplanes, and can be rescaled to
either $1, 0$ or $-1$. For a positive, zero or negative value of
each $\kappa_i$ the corresponding measure is respectively elliptic, parabolic
or
hyperbolic (hence the number $3^N$ of essentially
different CK systems). Alternatively, these coefficients can be
related to the  constant curvatures of the canonical connections on
the symmetrical homogeneous spaces of points, lines, \dots
hyperplanes. A complete description will be given in a forthcoming
paper \cite{\HOS}, but we only remark here that when $\k_1=1,0,-1$
and $\k_2=\k_3=\dots=\k_N=1$, then the CK space of points
$G/H^{(0)}$ is identified to the standard sphere $\Bbb{S}^N$, to the
euclidean space $\Bbb E^N$ or to the hyperbolic space $\Bbb H^N$ as
symmetric homogeneous spaces of constant curvature $\k_1$
corresponding to the groups (1.1) which are in this case either
$SO(N+1)$, $E(N)\equiv ISO(N)$ and $SO(N,1)$. Therefore, these three
well known \it riemannian \rm  symmetric spaces can be thought of as
a particular cases of the CK scheme, where the distance between
points is either elliptic, parabolic or hyperbolic, while all others
distances (between lines, \dots , hyperplanes) are elliptic.

\medskip

The aim of this paper is two--fold. First, we define $\Bbb Z_2^ {\otimes
N}$ graded contractions of the real algebras $so(N+1)$ for arbitrary
$N$. Second, we show how the $N$-dimensional CK algebras appear as a
distinguished family of these graded contractions of $so(N+1)$. The
order of $\Bbb Z_2^{\otimes N}$ is larger than the dimension of
$so(N+1)$, so that the number of irrelevant contraction parameters grows
rapidly as $N$ increases. The strategy of first solving the contraction
equations for a $2^N\times 2^N$ symmetric contraction ``universal"
matrix and then disregarding the elements which are irrelevant for the
specific case under discussion is perhaps not the best choice.
Therefore, we first identify all the relevant parameters for this grading
of $so(N+1)$, which fall into three classes, and only then we write down
in an adequate form the relevant contraction equations which are
completely solved for a special case. Thus the place occupied by the CK
algebras within the family of all graded contractions of $so(N+1)$ can
be appreciated very easily.

\medskip

The paper is organized as follows: next section presents a
brief overview about graded contractions and of the particular $\Bbb
Z_2^{\otimes N}$ grading we are using for $so(N+1)$. We classify the
relevant contraction parameters, and we write the contraction equations.
In Section III,  we show how the CK algebras appear naturally as a
subset of the graded contractions of $so(N+1)$, and we briefly
touch upon physical applications.

\bigskip
\bigskip

\centerline {\bf II. $\Bbb Z_2^{\otimes N}$ GRADED CONTRACTIONS OF so(N+1).}
\medskip
\medskip

\noindent
\subheading {II.1 Graded contractions}
\medskip

Let us recall briefly the theory of graded contractions of Lie algebras.
Suppose $L$ is a real Lie algebra, graded by an Abelian finite group
$\varGamma$ whose product is denoted additively. The grading is a
decomposition of the vector space structure of $L$ as
$$
L=\bigoplus_{\mu\in \varGamma} L_\mu ,  \tag2.1
$$
such that for $x \in L_\mu$ and $y \in L_\nu$, if $[x,y]$ is \it
non--zero, \rm then it belongs to $L_{\mu+\nu}$. This is written
symbolically as:
$$
0 \neq [L_\mu,L_\nu]\subseteq L_{\mu+\nu} , \qquad
\mu, \, \nu, \, \mu+\nu\in \varGamma.  \tag2.2
$$

A \it graded contraction \rm of the Lie algebra $L$ is a Lie
algebra $L_\ee$  with the same vector space structure as $L$, but Lie
brackets for $x\in L_\mu$, $y\in L_\nu$ modified as:
$$
[x,y]_\ee := \ee_{\mu,\nu} \,[x,y], \quad \text{in shorthand form}
\quad [ L_\mu, L_\nu ]_\ee := \ee_{\mu,\nu} \, [ L_\mu, L_\nu], \tag
2.3
$$
where the \it contraction parameters \rm $\ee_{\mu,\nu}$ are real
numbers such that $L_\ee$ is indeed a Lie algebra \cite{\MonPat}. From
antisymmetry and Jacobi identities, one easily gets the \it contraction
equations:\rm
$$
\align
&\ee_{\mu,\nu} = \ee_{\nu,\mu} \tag2.4a \\
&\ee_{\mu,\nu} \, \ee_{\mu+\nu,\sigma}=
\ee_{\mu,\nu+\sigma} \, \ee_{\nu,\sigma} \tag2.4b
\endalign
$$
for all \it relevant \rm values of indices. Condition (2.4a) means that
$\ee_{\mu,\nu}$ can be looked at as a symmetric matrix (the \it
contraction matrix, \rm see \cite{\MonPat ,\MooPat} for more details).
Each set of parameters $\ee$ which is a solution of (2.4) defines a
contraction; two contractions $\ee^{(1)}$, $\ee^{(2)}$ are equivalent if
they are related by:
$$
\ee^{(2)}_{\mu,\nu}=\ee^{(1)}_{\mu,\nu} \, \frac{r_\mu
r_\nu}{r_{\mu+\nu}},  \tag 2.5
$$
(without summation over repeated indices) where the $r$'s are \it
non--zero \rm real numbers which should be thought of as \it scaling
factors \rm of the grading subspaces in the Lie algebra.

Even if the contraction parameters associated to \it any \rm pair of
elements $\mu,\nu$ in $\varGamma$ seem to appear in the equations (2.4),
many of them will not, for two reasons:
  \roster
  \item In the direct sum  (2.1) only those $L_\mu$ which are proper
subspaces must be considered; the set of grading group elements $\mu$
actually appearing in the direct sum (2.1) is some subset of
$\varGamma$. The $\ee$'s containing an index $\mu$ outside this
subset will not appear in the system (2.4).
  \item It could happen that in the non--contracted algebra, all the
elements $x \in L_\mu$ commute with the elements $y \in L_\nu$; this
situation will be denoted symbolically as $[L_\mu, L_\nu]=0$. Of course,
the parameters $\ee_{\mu,\nu}$ corresponding to $[L_\mu, L_\nu]=0$ are
also completely irrelevant and the equations (2.4b) which contain such
parameters do not appear.
  \endroster
More details about graded contractions of Lie algebras are given in
\cite{\MonPat,\BerMon}. The analogous theory for the representations is
described in \cite{\MooPat}.

\medskip
\medskip
\noindent
\subheading {II.2 A fine grading for so(N+1)}
\medskip

In this work we consider the family of graded contractions of $so(N+1)$
which preserve a $\Bbb Z_2^{\otimes N}$ fine grading. The algebra
$so(N+1)$ has $N(N+1)/2$ generators $J_{ab},$ with $a<b; (a,b=0,1,\dots,
N)$. The non--zero Lie brackets are:
$$
a<b<c, \qquad \cases
[J_{ab}, J_{ac}] = J_{bc}, \\
[J_{ab}, J_{bc}] = -J_{ac}, \\
[J_{ac}, J_{bc}] = J_{ab}, \\
\endcases \tag 2.6
$$
(all Lie brackets involving four \it different \rm indices $a,b,a',b'$
like $[J_{ab}, J_{a'b'}]$ are equal to zero). The standard $(N+1) \times
(N+1)$ matrix realization is:
$$
J_{ab} = - E_{ab} + E_{ba}, \qquad a<b, \tag 2.7
$$
where $E_{ab}$ is the matrix with a single 1 entry at row $a$, column
$b$, and zeros at the remaining entries. Throughout all the paper we
will use $a,b,c,d$ as indices whenever these are implicitly assumed to
appear ordered (as in the generators $J_{ab}$), and
$i,j,k,l,m$ where no ordering is implied.

Let $\calI$ be the set of indices $\{0,1,\dots ,N\}$. We denote by
$\calS$ any subset of $\calI$ and $\chi_{\calS}(i)$ the characteristic
function over $\calS$:
$$
\chi_{\calS}(i) = \left\{
  \aligned 1 & \quad\text{\ if}\ i\in \calS,\\
           0 & \quad\text{\ if}\ i\notin \calS.
  \endaligned \right.  \tag2.8
$$

We define a linear mapping $S_\calS : so(N+1) \to so(N+1)$, associated
to $\calS$ as:
$$
S_{\calS} J_{ab} = (-1)^{\chi_{\calS}(a)+\chi_{\calS}(b)}J_{ab}.
\tag2.9
$$
The properties of $S_\calS$ are:
 \roster
   \item $S_\calS$ is an involutive automorphism (i.e. it provides a
$\Bbb Z_2$ grading) of $so(N+1)$.
   \item $S_{\calS} = S_{\calI\backslash\calS}$ (i.e. the automorphism
associated to a subset $\calS\subseteq \calI$ is the same as the one
associated to its complement $\calI\backslash\calS$ in the whole set of
indices $\calI$).
   \item For any two subsets $\calS, \calS'$ of $\calI$, we have
$S_\calS \cdot S_{\calS'} = S_{\calS \cup \calS'} \cdot S_{\calS
\cap \calS'} = S_{\Delta(\calS, \calS')}$, where $\Delta(\calS,
\calS')$ is the symmetric difference of the subsets $\calS, \calS'$.
 \endroster
For instance, if $\calI =\{0,1,2,3\}$, we have $
S_{12}=S_{03}=S_{0}S_{3}=S_{01}S_{13}=S_{013} S_{1}=\dots $, etc.

\medskip

{}From property (1) it follows that the generators $J_{ab}$, with either
both indices or none at all in $\calS$, span the $S_\calS$-invariant
subspace  of the Lie algebra $L\equiv so(N+1)$ whereas the antiinvariant
generators (i.e. those multiplied by $-1$) are the $J_{ab}$ with exactly
\it one \rm index in $\calS$. From (2), the total number of $S_\calS$ is
reduced from the number of subsets of $\calI$ to $2^N$ different
involutions. Finally, (3) shows that all these involutions commute, so
that they constitute an abelian group $\varGamma$, isomorphic to $\Bbb
Z_2^{\otimes N}$. This group is generated by $N$ involutions, for
instance,
$$
S_0, \quad S_{01},\quad S_{012}, \quad \dots \quad S_{0\dots N-1}.
\tag 2.10
$$
This particular set of involutions will play a special role, and
sometimes we denote them simply as $S^{(k)} \equiv S_{0\dots k},\
k=0,\dots N-1$.

\medskip

It is well known \cite{\PatZas} that each set of commuting automorphisms
of $L$  determines a grading, so $\varGamma$ becomes a grading group of
$so(N+1)$. A generic element $\mu \in \varGamma$ can be written as a
product of powers $\prod_{k=0}^{N-1} (S^{(k)})^{\mu_k}$, where $\mu_k
\in \{0,1\}$. Relatively to the system of generators (2.10) of
$\varGamma$, $\mu$ can be described by a string $\{ \mu_k \} \equiv
\{\mu_0 \mu_1 \dots \mu_{N-1} \}$ of $N$ elements containing $0$'s and
$1$'s. Not all $\mu \in \varGamma$ are associated to a non--empty
subspace $L_{\mu}$. Using (2.9) one can verify that the basis element
$J_{ab}, \, a<b$ belongs to the grading subspace $L_\mu$:
$$
\langle J_{ab} \rangle = L_\mu \quad \equiv \quad \mu= \{ {0\dots
0\underset {\ssize a}\to1\dots 1\underset {\ssize b}\to0\dots 0}
\}, \tag 2.11
$$
where all the $1$'s are in a contiguous string, starting at the $a$-th
position and ending at the $(b-1)$-th position, eventually preceded
and/or followed by strings of $0$'s. Therefore, out of the $2^N$
elements $\mu\in\varGamma$, only those $N(N+1)/2$ elements of the form
(2.11) have actual associated grading subspaces $L_\mu$. All subspaces
$L_\mu$ are one-dimensional and thus this  $\Bbb Z_2^{\otimes N}$
grading is fine \cite{\PatZas}.
\medskip
\medskip

\noindent
\subheading {II.3 The essential contraction parameters for the
fine grading of so(N+1)}
\medskip

Instead of using the complete string $\{\mu_k\}$ to describe $\mu$ when
solving equations  (2.4) it will prove very useful to denote the
particular $\mu$ (2.11) by the pair of indices $\mu \equiv ab, \
(a<b)$, and to speak of $a,b$ as the \it indices \rm of $\mu$. In the
following, let $\mu=ab$, $\nu=a'b'$.

The complete string associated to $\mu+\nu$ is:
 \roster
   \item A string with all zeros only if $\mu=\nu$ (i.e., $a=a'$
and $b=b'$; $\mu$ and $\nu$ have \it both \rm indices in common).
   \item A string with a single contiguous substring of $1$'s if
$\mu$ and $\nu$ have a \it single \rm index in common.
   \item A string with two substrings of $1$'s, separated
by $0$'s, if $\mu$ and $\nu$ have no indices in common.
 \endroster

Cases (1) and (3) apparently produce grading group elements $\mu+\nu$
without associated grading subspaces. However, the commutation relations
(2.6) shows that in cases (1) and (3), and only in these cases we have
identically $[L_\mu, L_\nu] = 0$. So, the only \it relevant \rm
contraction coefficients $\ee_{\mu,\nu}$ are those with $\mu$, $\nu$ of
the form (2.11) and with a \it single \rm index in common. They fall
naturally into \it three \rm  disjoint subsets,
$$
\ee_{ab,ab'} \qquad \ee_{ab,bc} =
\ee_{bc,ab} \qquad \ee_{ac,a'c}.  \tag 2.12
$$
By using the symmetry $\ee_{\mu,\nu}=\ee_{\nu,\mu}$ all these $\ee$ can
be expressed in terms of the three sets of relevant \it essential
\rm contraction parameters:
$$
\alpha_{a;bc} \equiv \ee_{ab,ac}, \qquad
\beta_{ac}^{\,\,b} \equiv \ee_{ab,bc}, \qquad
\gamma_{ab;c} \equiv \ee_{ac,bc}, \qquad a<b<c. \tag 2.13
$$
We sum up by classifying the contraction parameters into disjoint
classes:
  \roster
    \item""{\bf Class 1.} Irrelevant parameters (that do not appear in
the equations (2.4)), which consists of all $\ee_{\mu,\nu}$ where at
least one of the complete strings of $\mu$, $\nu$ or $\mu+\nu$ is not of
the form (2.11) (a single contiguous string of $1$'s).
\smallskip

    \item"" {{\bf Class 2.} Relevant parameters, which are those

$\ee_{\mu,\nu}$ where $\mu$, $\nu$ and $\mu+\nu$ have complete strings
of the form (2.11) with a single string of $1$'s. This class
could be naturally splitted into three subclasses:
      \itemitem{} {\bf Class 2a.} Elements $\ee_{ab,ab'}$, essential
elements $\aa{a;bc},$
      \itemitem{} {\bf Class 2b.} Elements $\ee_{ab,bc}=\ee_{bc,ab}$,
essential elements $\bb{ac}{b},$
      \itemitem{} {\bf Class 2c.} Elements $\ee_{ac,a'c}$, essential
elements $\cc{ab;c}$.}
   \endroster

The total number of (relevant and not relevant) contraction parameters,
taking into account only the symmetry (2.4a) is $2^N(2^N+1)/2$, but they
are reduced beforehand to $3 {{N+1}\choose{3}}$ essential relevant
parameters upon which the contraction equations (2.4) have to introduce
further relations.
\medskip
\medskip

\noindent
\subheading {II.4 The contraction equations}
\medskip

Whenever one considers contractions of Lie algebras or of their
representations, the major problem is to solve the contraction defining
equations, often consisting in very large systems quite tedious to be
solved by hand. A computer program was devised to treat these cases
\cite{\BerMon}. However, in spite of the fact that the grading group
$\Bbb Z_2^{\otimes N}$ is comparatively large, here we find it much more
illuminating to analyse these equations in terms of the splitting of the
relevant coefficients into the three classes (2a--2c); the results are
simple and transparent enough to justify this approach.

So we set to solve the equations (2.4b):
$$
\ee_{\mu,\nu} \, \ee_{\mu+\nu,\sigma} =
\ee_{\mu,\nu+\sigma} \, \ee_{\nu,\sigma}, \tag 2.14
$$
for all values of the three indices $\mu,\nu,\sigma$ giving rise to
relevant (Class 2) $\ee$'s. An elementary analysis shows that for fixed
$\mu$, $\nu$, which together determine a unique set of \it three \rm
different indices $i,k,l$ (see the comment above (2.12)), the only
possibilities for $\sigma$ leading to relevant $\ee$'s in (2.14) are
either $\sigma=\mu$ (in which case (2.14) reduces to an identity) or the
indices of $\sigma$ are the index in $\nu$ but not in $\mu$, and a new
fourth index $m$, different from $i, k, l$. So for each set of \it four
\rm different indices, $i,k,l,m$, there is a relevant equation of the
kind (2.14):
$$
\eee{ik}{il} \, \eee{kl}{lm} = \eee{ik}{im} \, \eee{il}{lm}, \qquad
i,k,l,m \text{ all different}, \tag 2.15
$$
where the arrow under each index pair means that these indices should be
always put in their natural order (so $\underset{\rightarrow}\to{ik}$
stands for $ik$ if $i<k$, but for $ki$ if $k<i$). Each of the $\ee$'s
appearing in this equation belong to the relevant classes 2a-2c.
Furthermore, (2.15) is invariant under the interchange of the first and
the second pairs of indices, $ik \lra lm$, so that out of the $4!=24$
equations associated to each set of four different indices $i,k,l,m$,
only \it twelve \rm equations are different.

\medskip

The next step through solving these equations is to rewrite them using
the natural ordering for the four different indices, say $a<b<c<d$, and
then replace all \it relevant \rm $\ee$'s by the \it relevant essential
\rm elements $\aa{a;bc}, \bb{ac}{b}, \cc{ab;c}$ with $a<b<c$
(see (2.13); e.g. $\ee_{bd,bc} = \ee_{bc,bd} = \aa{b;cd}, \,
\ee_{cd,bc} = \ee_{bc,\, cd} = \bb{bd}{c}$). The twelve equations
are:
$$
\alignat3
& &\quad \bb{ac}{b} \bb{ad}{c} &= \bb{ad}{b} \bb{bd}{c},
&\quad & \qquad\tag 2.16a\\
\aa{a;bc} \bb{bd}{c} &= \aa{a;bd} \bb{ad}{c}, &\quad  \aa{a;bd}
\aa{b;cd} &= \aa{a;cd} \bb{ac}{b}, &\quad  \aa{a;bc} \aa{b;cd} &=
\aa{a;cd} \bb{ad}{b} \qquad\tag 2.16b\\
\aa{a;cd} \cc{bc;d} &= \aa{a;bc} \cc{ab;d}, &\quad & &\quad
\aa{a;bd} \cc{bc;d} &= \aa{a;bc} \cc{ac;d} \qquad\tag 2.16c\\
\bb{ad}{b} \cc{ac;d} &= \bb{ac}{b} \cc{bc;d}, &\quad
\bb{bd}{c} \cc{ab;d} &= \cc{ab;c} \cc{ac;d}, &\quad
\bb{ad}{c} \cc{ab;d} &= \cc{ab;c} \cc{bc,d} \qquad\tag 2.16d\\
\aa{a;cd} \bb{bd}{c} &= \aa{a;bd} \cc{ab;c}, &\quad
\aa{b;cd} \bb{ad}{c} &= \bb{ad}{b} \cc{ab;c}, &\quad
\aa{b;cd} \cc{ac;d} &= \bb{ac}{b} \cc{ab;d}\qquad  \tag 2.16e
\endalignat
$$
There is a single equation involving \it only \rm $\beta$'s, while
all others relate $\ee$ parameters belonging to either two or the
three subclasses 2a-2c. This fact, which is emphasized in the
presentation of the set (2.16) would be difficult to see by solving
the equations for $\ee_{\mu,\nu}$ by means of a computer program,
because the special role played by the contraction parameters
$\ee_{ab,bc}$ would not be highlighted.

It is therefore clear that the strategy for solving the set of
equations (2.16) is first to start with  (2.16a), and for each set of
$\beta$'s solving it, substitute into the remaining equations which
are then considerably simplified. It is not our goal here to provide
the general solution of the complete system, but rather to see how
the ``simplest" solutions lead exactly to the CK algebras.

\bigskip
\bigskip

\centerline {\bf III. CK ALGEBRAS AS $\Bbb Z_2^{\otimes N}$ CONTRACTIONS
OF so(N+1).}
\medskip
\medskip

\noindent
\subheading {III.1 Quasi--regular solutions to the grading
equations}
\medskip

In the general case, the solutions of the contraction equations are
classed into ``regular" (all $\ee$'s different from zero) and
``non--regular" ones (some $\ee$'s equal to zero). Here we refine this
classification and call ``quasi--regular" the
solutions with all  $\bb{ac}{b} \neq 0$. The ``simplest"
quasi--regular solution of (2.16a) has all $\bb{ac}{b} = 1, \
a<b<c$.  (We shall see in paragraph III.2 that any contraction with
\it all \rm $\bb{ac}{b} \neq 0$ is equivalent to a contraction with
all $\bb{ac}{b} = 1$, so these are essentially the only
quasi--regular solutions.) Fixing $\beta =1$, equations (2.16)
reduces to $$
\alignat2
\aa{a;bc} &= \aa{a;bd}, \qquad\qquad
\aa{b;cd} = \cc{ab;c}, &\quad
\cc{ac;d} &= \cc{bc;d}, \qquad \tag3.1a \\
\aa{a;cd} &=\aa{a;bd} \aa{b;cd} = \aa{a;bc} \aa{b;cd},
\ &\quad
\cc{ab;d}=\cc{ab;c} \cc{ac;d} &= \cc{ab;c} \cc{bc;d},  \qquad\tag3.1b\\
\aa{a;cd} &= \aa{a;bd} \cc{ab;c},
\  &\  \
\cc{ab;d} &= \aa{b;cd} \cc{ac;d},  \qquad\tag3.1c \\
\aa{a;cd} \cc{bc;d} &= \aa{a;bc} \cc{ab;d},
\  &\  \
\aa{a;bd} \cc{bc;d} &= \aa{a;bc} \cc{ac;d},  \qquad\tag3.1d\\
\endalignat
$$
where $a<b<c<d$.

\medskip

The first equation in (3.1a) indicates that as long as $b<c$,
$\aa{a;bc}$ does not depend on $c$; we may denote them as a two--index
object, $A_{ab} \equiv \aa{a;bc}$. Likewise, the third equation in
(3.1a) implies that $\cc{bc;d}$ is actually independent of $b$ as long
as $b<c$, so we may denote it as $C_{cd} \equiv \cc{bc;d}$. The
remaining equation in the group (3.1a) implies $A_{bc} = C_{bc}$.
Replacing $\aa{b;cd} = \cc{ab;c} = A_{bc}$ in (3.1b-d), all these
equations reduce to a \it single \rm independent equation for each set
of three ordered indices:
$$
A_{ac} = A_{ab} A_{bc},\qquad a<b<c. \tag 3.2
$$
If we call the \it index difference \rm of the
object $A_{ab}$, the \it positive \rm integer $b-a$, then
equations (3.2) express those $A_{ac}$ with a given index
difference in terms of those with a \it smaller \rm index
difference. This means that ultimately only those $A_{ab}$ with
index difference equal to $1$, $A_a := A_{a-1,a}, \  a=1,2,\dots
N$ will be independent. All other $A_{ac}$  are expressed in terms
of $A_a$ as:
$$
A_{ac} = A_{a+1} A_{a+2} \dots A_c. \tag 3.3
$$

So, we have obtained the:

\proclaim {\bf Theorem 1} Within the Ansatz $\bb{ac}{b}=1$, any
solution of the contraction equations (2.16) is determined by a set
of $N$ independent real numbers, $A_a$, $a=1, \dots N$, and is given by:
$$  \align
{\text{Class 2a }}\qquad &\ee_{ab,ac} \equiv \aa{a;bc} = A_{a+1}
\dots A_b, \\
{\text{Class 2b }}\qquad &\ee_{ab,bc} \equiv \bb{ac}{b} = 1, \tag
3.4\\ {\text{Class 2c }}\qquad &\ee_{ac,bc} \equiv \cc{ab;c} =
A_{b+1} \dots A_c. \endalign
$$
\endproclaim

Those contractions with all $A_a \geq 0$ belong to the continuous type,
all the others being of discrete type (the continuous or discrete
character of a contraction is defined in \cite{\MonPat}). The  graded
contractions of $so(N+1)$ with all $A_a \neq 0$ and at least one $A_a <
0$ provide the different non--compact real forms $so(p,q)$, with
$p+q=N+1$, of $so(N+1)$.

\medskip
\medskip

\noindent
\subheading {III.2 Equivalence between quasi--regular solutions}
\medskip

Let us return now to equations (2.16). Under a general scale change,
$J_{ab} \to r_{ab}\, J_{ab}$,  (no sum in repeated indices!) with
$r_{ab} \neq 0$, the essential relevant contraction coefficients change
as:
$$
\aa{a;bc} \to
\aa{a;bc}  \frac{r_{ab} r_{ac}}{r_{bc}}, \qquad
\bb{ac}{b} \to
\bb{ac}{b} \frac{r_{ab} r_{bc}}{r_{ac}}, \qquad
\cc{ab;c} \to
\cc{ab;c}\frac{r_{ac} r_{bc}}{r_{ab}}. \tag 3.5
$$
 Now we set to use the freedom allowed by these changes to
reduce each possible solution of the contraction equations to some
``standard" form.

\medskip

We start with the $\beta$'s. Let us
call the \it index difference \rm of $\bb{ac}{b}$ the \it positive \rm
integer $c-a$. Due to the presence of a third index $b$ (such that
$a<b<c$) the index difference of any $\bb{ac}{b}$ is always greater than
or equal to 2, and the number of different $\bb{ac}{b}$ with the same
pair $ac$ is one less than their index difference.

\medskip

Consider the elements $\bb{a,a+2}{a+1}$, whose index difference is equal
to 2. Each of them can be reduced to $1$ (as long as they are different
from zero) by adjusting the scale coefficients with the same index
difference $r_{a,a+2}$ (i.e. take $r_{a,a+2} = \bb{a,a+2}{a+1} \,
r_{a,a+1}\, r_{a+1,a+2}$, so that according to (3.5) the new
$\bb{a,a+2}{a+1} \to 1$).

\medskip

Now consider the elements $\bb{a,a+3}{a+1}$, whose index difference is
equal to 3. By using the same procedure, each of them can be reduced to
$1$ (as long as they are different from zero) by adjusting the scale
coefficients $r_{a,a+3}$ (i.e. take $r_{a,a+3} = \bb{a,a+3}{a+1} \,
r_{a,a+1}\, r_{a+1,a+3}$, so that the new $\bb{a,a+3}{a+1} \to 1$). By
iterating, it is clear that this procedure reduces to 1 all those
contraction coefficients $\bb{a\ c}{a+1}$, as long as they are
non--zero. But the remaining $\beta$'s are not independent, and should
still satisfy equations (2.16a). In particular, if all $\bb{a\
d}{a+1}=1$, these equations for $a,a+1,a+2$ and $d$ imply:
$$
\bb{a\,,\, a+2}{a+1} \bb{a\,,\, d}{a+2} = \bb{a\,,\, d}{a+1}
\bb{a+1\,,\, d}{a+2},
$$
so that $\bb{a\ d}{a+2}=1$. The same procedure with $a,a+2, a+3, d$ now
implies $\bb{a\ d}{a+3}=1$, and so on. Thus, once the $\bb{a\ c}{a+1}$
are equal to 1, \it all \rm $\beta$'s are equal to 1. This justifies the
statement made at the beginning of paragraph III.1.

\medskip

Now, it remains to exploit the freedom still left to reduce the
remaining constants $A_a$ to some standard values. This should be done
without spoiling the equalities $\bb{ac}{b}=1$. By (3.5), any further
scale change with $r_{ac} = r_{ab} r_{bc}, \ a<b<c$ will keep unchanged
the values of \it all \rm $\beta$'s. This means that the scale parameters
$r_{a,a+1}$ (with index difference equal to 1) are the only free scaling
coefficients, all the others being determined through the relation:
$$
r_{ab} = r_{a,a+1} r_{a+1,a+2} \dots r_{b-1,b}. \tag 3.6
$$
The behaviour of the remaining essential relevant contraction
coefficients $\aa{b;cd} = \cc{ab;c}$ under such a scale change is
$$
\aa{b;cd} \to \aa{b;cd} \, \frac{r_{bc} r_{bd}}{r_{cd}} =
\aa{b;cd} \, \frac{r_{bc} r_{bc} r_{cd}}{r_{cd}} =
\aa{b;cd} \, (r_{bc})^2, \qquad a<b<c, \tag 3.7
$$
and in particular, the change for $A_a \equiv A_{a-1,a} = \aa{a-1;ab}$ is
$$
A_a \to A_a \, (r_{a-1,a})^2, \qquad a=1, \dots N. \tag 3.8
$$
As there are $N$ free scaling coefficients $r_{a-1,a}$, and $N$
essential contraction parameters $A_a$ for the solution (3.4), it is
clear that each $A_a$ can be reduced to the standard values $A_a\in \{1,
0, -1\}$.

\medskip

We can sum up the results obtained in the previous paragraphs as:

\proclaim{\bf Theorem 2} Any quasi--regular solution to the
contraction equations (2.16) can be reduced by equivalence to
one of the $3^N$ particular solutions determined in Theorem 1 by a
family of $N$ values $A_a, a=1, \dots N$, each $A_a$ taking values
in the set $\{1,0,-1\}$.
\endproclaim

\medskip
\medskip

\noindent
\subheading {III.3 The CK algebras as the quasi--regular contracted
algebras of so(N+1)}
\medskip

Once the pertinent solutions to the contraction equations have been
determined, the contracted Lie algebra obtained from $so(N+1)$ (see (2.6)) with
the
contraction parameters $\ee$ given in (3.4) has the following non--zero
Lie brackets:
$$
a<b<c, \qquad \cases
[J_{ab}, J_{ac}] = \k_{ab} J_{bc}, \\
[J_{ab}, J_{bc}] = -J_{ac},  \qquad
 \k_{ab}= \prod_{i=a+1}^b\k_i,\qquad a,b=0,1,\dots,N; \\
[J_{ac}, J_{bc}] = \k_{bc} J_{ab}, \\
\endcases \tag 3.9
$$
(all Lie brackets involving four \it different \rm indices $a,b,a',b'$
as $[J_{ab}, J_{a'b'}]$ are again equal to zero). It is clear that (3.9)
coincides with
(1.1), so that the family of quasi--regular graded contractions of
$so(N+1)$ leads exactly to the family of CK algebras, and the
contraction parameters $A_a$ are to be identified with the
geometrical constants $\k_a$ appearing in the CK scheme. The
commutation relations (3.9) include the Lie algebras of $SO(p,q),\
(p+q=N+1)$, when all $\k_a \ne 0$, those of the inhomogeneous
$ISO(p,q),\ (p+q=N)$ when, for instance, $\k_1=0$ but all other
$\k_a\ne 0$; as well as many other different algebras when more
$\k_a$ are equal to zero. It is a satisfying and unifying result to
find that all of these groups, first studied in connection with
projective geometry and then with projective metrics,
come out as the more ``regular" family of graded contractions of
$so(N+1)$. \medskip

It is interesting to inquire on the geometrical meaning of
the In\"on\"u--Wigner (IW) type contractions associated to each of the
$\Bbb Z_2$ subgradings of our grading. Each of these contractions is
associated to an involution $S_\calS$. The effect of the IW contraction
is to perform a graded scale change (by a factor $\lambda$) on the
generators multiplied by $-1$ under $S_\calS$ (i.e. those with a single
index in $\calS$) and then to take the limit $\lambda \to 0$. The effect
of this scale change on the parameters $\aa{i;jk}, \bb{jk}{i}$ and
$\cc{jk;i}$ can be described simply as:

 \roster
   \item If either both sets $\{i\}$ and $\{jk\}$ or none at all are
contained in $\calS$, then the parameters $\aa{i;jk}, \bb{jk}{i}$
and $\cc{jk;i}$ do not change.
   \item If only one of these sets is contained in $\calS$, then the
parameters $\aa{i;jk}, \bb{jk}{i}$ and $\cc{jk;i}$ inherit a
$\lambda^2$ factor (so they vanish in the limit $\lambda \to 0$).
 \endroster

In particular, the IW contractions related to the involutions
$S_\calS \equiv S^{(M)}$ with the
particular subsets $\calS = \{0, 1, \dots , M\}$ are the
contractions around a point in $G/H^{(0)}$ (for $M=0$), a
line in $G/H^{(1)}$ (for $M=1$), \dots , a hyperplane in
$G/H^{(N-1)}$ (for $M=N-1$). Due to the
ordered nature of the indices in $\alpha, \beta, \gamma$, it is
clear that for these particular contractions the $\beta$'s do not
change. The effect on the  parameters $A_a$ is to make $A_{M+1}$
equal to zero, while keeping unchanged the remaining contraction
parameters, in complete accordance with the geometrical
interpretation. The
contractions associated to another type of subset of $\calI$ (for
instance $\calS = \{02\}$) would describe the behaviour of a CK
geometry near another kind of line, which may be of a different type
than the standard one. For instance, a contraction of Poincar\'e
group around a time--like line in Minkowski space leads to the
Galilei group, while a contraction around a space--like line (which
is of different kind) leads to the Carroll group and to a different
CK geometry.

But these IW contractions do not exhaust all the
possible relations within the CK scheme. For instance, sequences of
IW contractions, which cannot be described as the limit of a scale
change with factors either $1$ or $\lambda$ are included in the set
of graded contractions.

\medskip

Let us now illustrate the result of Theorem 2 by some examples. We first
consider the 2-dimensional case. The basis is $\{J_{01}, J_{02},
J_{12}\}$, with commutation relations
$$
[J_{01},J_{02}]=\kappa_1 J_{12},\quad
[J_{01},J_{12}]=-J_{02},\quad [J_{02},J_{12}]=\kappa_2
J_{01} .\quad  \tag 3.10
$$
Since $N=2$, the ensuing CK algebras have a $\Bbb Z_2\otimes\Bbb Z_2$
grading generated  by the automorphisms $S^{(0)}$ and $S^{(1)}$ which
act as (from (2.9))
$$
\aligned
&S^{(0)} \equiv S_0: (J_{01},J_{02},J_{12}) \longrightarrow
(-J_{01},-J_{02},J_{12}),\\
&S^{(1)} \equiv S_{01}: (J_{01},J_{02},J_{12}) \longrightarrow
(J_{01},-J_{02},-J_{12}).
\endaligned \tag3.11
$$

These involutions provide the $so(3)$ basis with the following grading
$$
L_{\{01\}} \equiv L_{12}= \langle J_{12}\rangle , \quad
L_{\{10\}} \equiv L_{01}= \langle J_{01}\rangle , \quad
L_{\{11\}} \equiv L_{02}= \langle J_{02}\rangle , \tag 3.12
$$
where the subscripts enclosed in braces (in $L_\mu$) denote the complete
string $\{\mu_k\}$, while the subscripts not in braces denote the
indices of the relevant $\mu$. The total number of relevant essential
contraction parameters is ${3} {N+1\choose3}=3$ (one $\alpha$, $\beta$,
$\gamma$ each). With $\beta=1$, the others are related to the constants
$\k_a$ as:
$$
\k_{12}=\k_2 \lra \ee_{\{01\},\{11\}} = \ee_{02,12} \equiv
\aa{01;2}, \qquad
\k_{01}=\k_1 \lra \ee_{\{10\},\{11\}} = \ee_{01,02} \equiv
\cc{0;12}. \tag 3.13
$$
In these and in subsequent expressions, the order of each pair of
subindices in $\ee$ has been always adjusted to either the lexicographic
order when $\mu$ is described by its complete string or to the order
corresponding to the essential relevant elements when $\mu$ is described
through its indices.

\medskip

Now we consider the 3-dimensional case, with basis
$$
\{J_{01}, J_{02}, J_{03}, J_{12}, J_{13}, J_{23}\}. \tag 3.14
$$

The $\Bbb Z_2 ^{\otimes 3}$ determines the grading subspaces:
$$ \aligned
L_{\{100\}} \equiv L_{01} = \langle J_{01} \rangle, \quad
L_{\{110\}} \equiv L_{02}&= \langle J_{02} \rangle, \quad
L_{\{111\}} \equiv L_{03} = \langle J_{03} \rangle, \quad \\
L_{\{010\}} \equiv L_{12} = \langle J_{12} \rangle, \quad
L_{\{011\}} \equiv L_{13}&= \langle J_{13} \rangle, \quad
L_{\{001\}} \equiv L_{23} = \langle J_{23} \rangle. \quad
\endaligned \tag 3.15
$$

In this case, out of the $\frac{2^N(2^N+1)}2 = \frac {8(8+1)}2=36$
contraction parameters only $3 {{N+1}\choose 3} = 3 {4\choose3} = 12$
are the essential relevant ones: the four $\beta$'s are restricted to
$1$:
$$
\aligned
1 \lra \ee_{\{010\},\{100\} } = \ee_{01,12} \equiv \bb{02}{1}, \qquad
1 \lra \ee_{\{001\},\{010\} } = \ee_{12,23} \equiv \bb{13}{2}, \\
1 \lra \ee_{\{011\},\{100\} } = \ee_{01,13} \equiv \bb{03}{1}, \qquad
1 \lra \ee_{\{001\},\{110\} } = \ee_{02,23} \equiv \bb{03}{2}, \\
\endaligned \tag 3.16
$$
and the four remaining $\alpha$'s and $\gamma$'s
are expressed in terms of the $A_a \equiv \k_a , a=1,2,3$:
$$
\alignedat 7
&\k_{01}=\k_1 & &\ \lra \ & \{&\ee_{\{100\},\{110\}} =
\ee_{\{100\},\{111\}}\} & &= & \{&\ee_{01,02}=\ee_{01,03} \} &\
&\equiv &\ \{ &\aa{0;12} =   \aa{0;13}\}, \\
&\k_{02}=\k_1\k_2 & &\ \lra \ &   &\ee_{\{110\},\{111\}}
& &= &   &\ee_{02,03}                &\ &\equiv &\    &\aa{0,23}, \\
&\k_{12}=\k_2     & &\ \lra \ & \{&\ee_{\{010\},\{011\}} =
\ee_{\{010\},\{110\}}\} & &= & \{&\ee_{12,13}=\ee_{02,12} \}
&\ &\equiv &\ \{ &\aa{1;23} =  \cc{01;2}\}, \\
&\k_{13}=\k_2\k_3 & &\ \lra \ &   &\ee_{\{011\},\{111\}} & &= &
&\ee_{03,13}                &\ &\equiv &\    &\cc{01;3},  \\
&\k_{23}=\k_3     & &\ \lra \ & \{&\ee_{\{001\},\{011\}} =
\ee_{\{001\},\{111\}}\} & &= & \{&\ee_{13,23}=\ee_{03,23} \}
&\ &\equiv &\ \{ &\cc{12;3} =  \cc{02;3}\}. \endalignedat\tag3.17
$$

For the contracted group we get the nonzero commutation
relations:
$$
\alignedat4
[J_{01}, J_{02}] &= \k_{01} J_{12}, &\quad
[J_{01}, J_{03}] &= \k_{01} J_{13}, &\quad
[J_{02}, J_{03}] &= \k_{02} J_{23}, &\quad
[J_{12}, J_{13}] &= \k_{12} J_{23},
\\
[J_{01}, J_{12}] & =      - J_{02}, &\quad
[J_{01}, J_{13}] &=       - J_{03}, &\quad
[J_{02}, J_{23}] &=       - J_{03}, &\quad
[J_{12}, J_{23}] &=       - J_{13},
\\
[J_{02}, J_{12}] & =\k_{12} J_{01}, &\quad
[J_{03}, J_{13}] &= \k_{13} J_{01}, &\quad
[J_{03}, J_{23}] &= \k_{23} J_{02}, &\quad
[J_{13}, J_{23}] &= \k_{23} J_{12}.
\endalignedat\tag 3.18
$$

\medskip

In summary, we have shown that the Lie algebras of the groups of
motions of the $N$--dimensional CK geometries  can be obtained as a
naturally distinguished subset of $\Bbb Z_2^{\otimes N}$ contractions
of the Lie algebra $so(N+1)$, according to the scheme described above.
The kinematical groups appear in this formalism as CK groups and they
  have been studied elsewhere both from the point
of view of $\Bbb Z_2 \otimes \Bbb Z_2$ graded contractions of $so(5)$
\cite{\MonPatTol}, and in relation with CK groups
and their quantum deformations \cite{\BHOSi,\BHOSii}. The
relationship with the present work can be established easily, by
considering the six possible  $\Bbb Z_2 \otimes \Bbb Z_2$
subgradings of the $\Bbb Z_2^{\otimes 4}$ grading we have considered
for the case $N=4$; only two are compatible with the kinematical
group requirements. We refer the interested reader to
\cite{\MonPatTol} for a more detailed account.

\medskip

A last question would be: Do the contraction parameters correspond
to physical parameters? There is indeed a direct connection in
some cases. In the usual non--relativistic limit, $1/c^2$ appears
as the contraction parameter $A_2$ associated to time--like lines,
while $A_1$ (associated to points) is to be identified to the
curvature of the DeSitter space--time (either $1/R^2$ or $-1/R^2$)
when contracting from the DeSitter to the Minkowski space. Similar
interpretations can be also made for other contraction parameters.

\bigskip

\subheading  {Acknowledgement} \medskip

We are indebted to Prof. R. C. Myers and to A. Ballesteros for
reading the manuscript, and to the referees for their pertinent
suggestions. M O thanks the CRM de l'Universit\'e de Montr\'eal,
where this work was started, for its hospitality, and  M M thanks
the Departamento de F\'\i sica Te\'orica de la Universidad de
Valladolid for its kindness. This work has been partially supported
by CICYT of Spain (project \# PB91/0196), by the program of
Formaci\'on de Personal Investigador  del Ministerio de Educaci\'on
y Ciencia of Spain (graduate grant of F J H), by the NSERC of Canada
and by the Fonds FCAR du Qu\'ebec (M M).

\end